\documentclass[fleqn,10pt]{article}
\usepackage[utf8]{inputenc}
\usepackage[T1]{fontenc}
\usepackage{authblk}
\usepackage[utf8]{inputenc}
\usepackage{graphicx}
\usepackage[margin=0.8in]{geometry}
\usepackage{xcolor}
\usepackage{amsmath,amssymb}
\usepackage{float}
\title{Fabrication of damage-free and/or contamination-free sub-$\mu$m electrodes using PMMA masks}

\author[1,2,3,*]{Alexey A. Kaverzin}
\author[3]{Bart J. van Wees}
\author[1,2,4]{Eiji Saitoh}

\affil[1]{Department of Applied Physics, The University of Tokyo, Tokyo 113-8656, Japan}
\affil[2]{Institute for AI and Beyond, The University of Tokyo, Tokyo 113-8656, Japan}
\affil[3]{Zernike Institute for Advanced Materials, University of Groningen, Groningen, 9747 AG, The Netherlands}
\affil[4]{WPI Advanced Institute for Materials Research, Tohoku University, Sendai 980-8577, Japan}
\affil[*]{akaverzin@g.ecc.u-tokyo.ac.jp}

\begin{document}

\maketitle

\begin{abstract}
Quality of the electrical contacts and interfaces in various metal/semiconductor/insulator heterostructures is one of the pivotal aspects in both applied and fundamental research areas. For instance, non-optimal contact resistance can limit the overall efficiency of a certain developed technology and thus considerably narrow the range or fully block its practical application. On the other hand in fundamental research it is often the case that the manifestation of targeted phenomenon crucially depends on the level of contamination in the fabricated experimental samples. Here we offer a set of recipes that are aimed at contamination-free and damage-free fabrication of the devices, mostly developed for the two dimensional materials, but nevertheless applicable for a wider range of the systems, where the quality of the interfaces and/or non-invasiveness of the fabrication recipes are important. Our recipes are based on the preparation of the flexible PMMA membranes, with the help of which we can prepare residue-free or damage-free electrical connections to the studied material.
\end{abstract}

\twocolumn
\sloppy

\section*{Introduction}
Quality of the contacts/interfaces between the constituent parts of any microelectronic device is undoubtedly essential for its reliability, reproducibility and overall efficiency. For example, in transistor-based devices preparing a clean low-resistance contact between the metallic leads and semiconductor channel is necessary for optimal operation speed and heat dissipation. In spin-transfer torque magnetic random-access memory (STT-RAM) devices imperfections of the interfaces between the ferromagnetic layers contribute to the scattering of the spin polarisation of the current, and, therefore, influence the performance of the memory unit. In spin-orbit torque RAM devices the interface plays even more essential role since the interface itself is the source of the spin orbit coupling which generates the spin obit torque and, therefore, its quality determines the efficiency of the device. 

For layered materials clean contacts and interfaces similarly play an essential role. For instance, it has been a long standing discrepancy between the experimental observations and theoretical predictions for the spin relaxation time in graphene-based lateral spin valves. Theoretically it was predicted to be $\simeq1\,\mu$s \cite{Pesin2012}, whereas experimentally it has been consistently measured to range from $~100\,$ps up to $~10\,$ns \cite{Han2014} depending on the substrate. One of the likely explanations of this discrepancy is the extrinsic scattering of the electron spin caused by the imperfections (pinholes) in the oxide layer separating graphene and deposited ferromagnetic metal\cite{Volmer2013}. There are experimental and theoretical reports [to be added] suggesting that clean van der Waals contact between the metallic electrodes and semiconducting layered materials provides an Ohmic linear-in-bias contact resistance compared to the Schottky barrier formation at the contact interface when metallic electrodes are deposited

All the examples given above serve to emphasise the critical importance of an appropriate quality of the contact interfaces in devices used for both applications and fundamental research. In this work we offer a full and detailed description of the fabrication methods that are targeted at the fabrication of residue free interfaces between the layered material and evaporated/transferred metals/oxides. These recipes have been developed over the recent years, with all of them being successfully employed for our previous research \cite{Ghiasi2019, Damerio2021, Hidding2021, Kaverzin_2022} yet without a focused description of the recipe in the published reports.

All the recipes employ the polymethyl methacrylate (PMMA) membranes as flexible shadow masks or carrier layers. One of the proposed methods allows the preparation of the contamination-free electrodes with dimensions as small as $\simeq100\,$nm (potentially it is limited by the resolution of the used mask and with appropriate optimisation it can be further reduced to $\simeq30\,$nm, Section~3) and imply that at all stages of the fabrication the target material does not get in touch with the polymer and does not get immersed into any of the solvents (both of them being a potential source of the contamination), which is otherwise unavoidable for the common lithography-based recipes. With another recipe we demonstrate how to prepare the metallic electrodes on a separate substrate and subsequently transfer them within the inert atmosphere onto the surface of the targeted material to form a van der Waals contact between the metallic electrodes and the studied material (Section~2). This fabrication method is of direct relevance for the air-sensitive and/or delicate materials that cannot withstand, for instance, direct spattering or evaporation of the metallic films. These recipes aim at fabrication of the composite devices with minimal level of contamination at the interfaces and with minimum damage from the fabrication procedure to the studied material, and therefore offer a clear advantage in comparison with the commonly used lithography-based techniques.

\section{Contamination-free contact interface and plasma etching}

\begin{figure}[h!]
\centering
\includegraphics[width={\linewidth}]{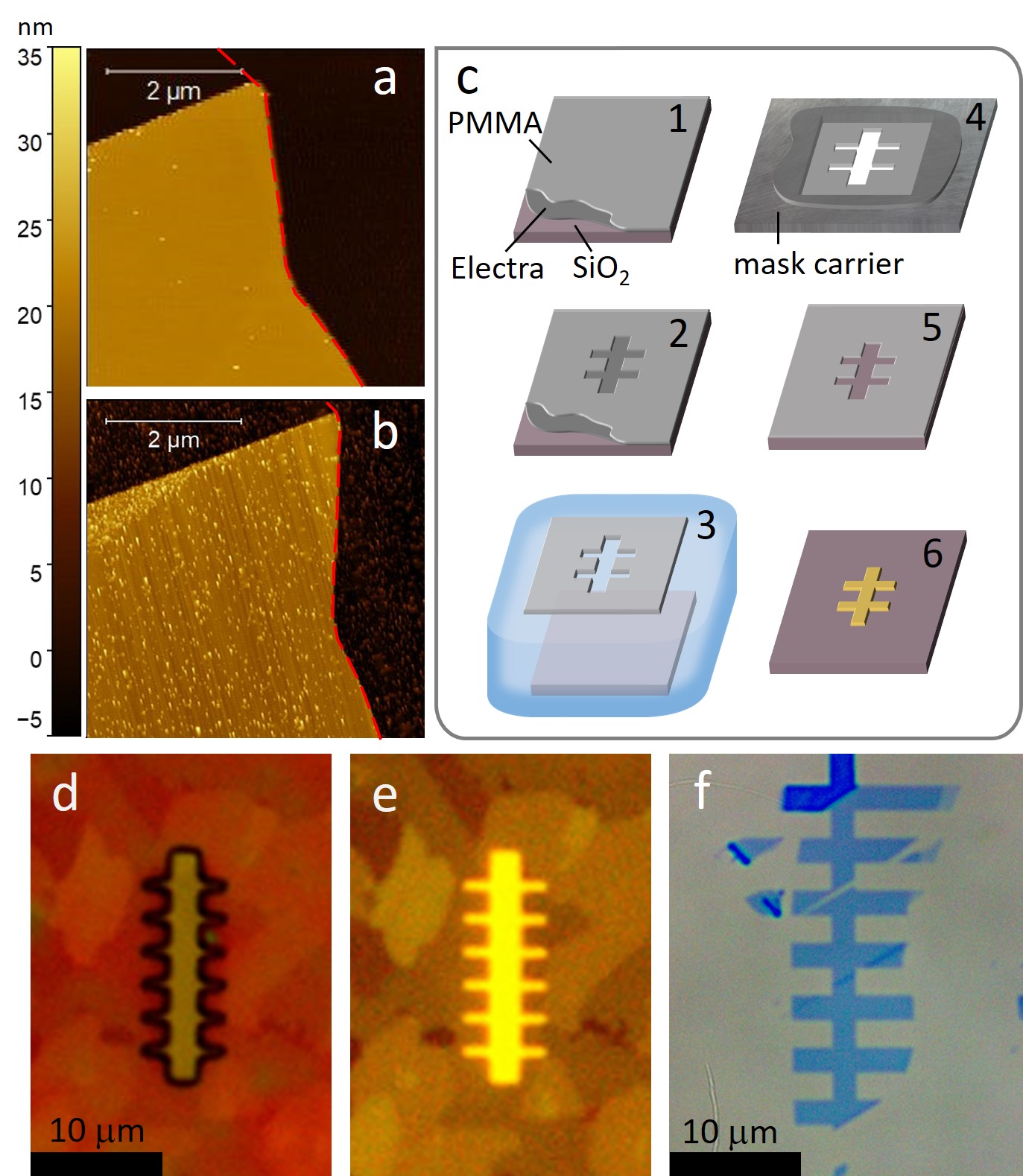}
\caption{AFM scan of the graphite flake after an e-beam lithography step performed using MMA-MAA copolymer (a) and 950K A4 PMMA (b). Red dashed line shows the edge of the graphite flake before etching. (c) The steps of the recipe are shown schematically and numbered from 1 to 6. (d) Optical image of the CaFe$_2$O$_4$ sample covered by the PMMA mask with a Hall-bar shaped opening, corresponding to step 5 in panel (c). (e) Optical image of the CaFe$_2$O$_4$ sample after deposition of Pt through the PMMA mask and mechanical removal of the mask, corresponding to step 6 in panel (c). (f) Optical image of the monolayer graphene flake after etching in the oxygen plasma with using PMMA membrane with rectangular openings as the etch mask. After etching the mask was removed mechanically before taking the shown image.}
\label{fig1}
\end{figure}

\subsection*{Surface contamination by PMMA}
Commonly used lithography-based fabrication of the electrodes includes the following steps: the spin-coating of a polymer resist layer on top of the target sample/substrate, exposure of the resist (either by e-beam or light), development of the exposed areas (assuming positive type resist), deposition of the metal films and, finally, lift off. During the spin-coating the full sample surface gets in direct contact with the resist. Subsequent immersion of the sample either in the developer or in the solvent removes the bulk of the resist, but a certain amount of it inevitably stays on the sample surface as contamination. Such contamination may not significantly influence the bulk properties of a sufficiently thick material, however, thin (layered) materials and interfaces are affected substantially \cite{Gammelgaard2014,Liang2019}. 

Fig.~1(a) demonstrates atomic force microscopy (AFM) scans of the graphite flake at different stages of the fabrication procedure. At first, after being exfoliated on the SiO$_2$/Si substrate, the graphite flake was spin-coated with MMA-MAA copolymer \cite{copolymer}. Subsequently, the resist was exposed with e-beam lithography and developed. Using reactive-ion plasma the open parts of the graphite flake were dry-etched. Next, by washing in acetone the remaining resist was removed and the AFM scan obtained (shown in panel (a)). As it is seen, a large portion of the graphite flake stays clean with estimated roughness of $\simeq0.4\,$nm (similar to that of a freshly exfoliated graphene/graphite on the SiO$_2$ surface) with some inclusion of rare contamination particles. Near the etched edge (top side of the flake on panels (a) and (b)) the contamination is more prominent, most likely due to the cross-linked during plasma etching parts of the copolymer, that could not be dissolved and removed by acetone. Then, the same graphite flake was spin-coated with PMMA (950K A4) and without any further lithography washed again in acetone. Panel (b) shows its surface scan directly after the final cleaning. Distinctly the level of surface contamination raised dramatically with increase of average roughness up to $\simeq2.4\,$nm.

Comparison between panels (a) and (b) of Fig.~1 evidently demonstrates that (1) covering the surface of the sample with resist polymer inevitably results in the increased contamination level of the surface (and up to an order of magnitude increase in surface roughness) and (2) the level of contamination depends on the presence of poorly soluble (long-chain polymer) particles. Thus, the conventional lithography-based fabrication of the metallic electrodes inevitably suffers from such polymer contamination, which affects the properties (such as electrical resistance) of the material itself and of the contact between the target material and deposited metals. In order to fully eliminate the possibility of such contamination we developed a recipe where the lithography of the resist (PMMA in our case) is performed on a separate sacrificial substrate as described in detail in the next subsection.

\subsection*{The recipe}
Fig.~2(c) summarises the steps of the fabrication and transfer of the PMMA mask used for preparation of contamination-free electrodes. In the first step (marked with "$1$"), a water-soluble polymer Electra \cite{Electra} is spin-coated on the surface of the substrate at the rate of $1000\,$rpm and subsequently baked at $90^{\circ}$C for $90\,$sec on the hot plate. Directly after, 950K A4 PMMA is spincoated at the rate of $1000\,$rpm and baked at $180^{\circ}$C for $90\,$sec. The resulting thicknesses estimated from the AFM and profilometer scans are $\simeq200\,$nm and $\simeq500\,$nm for the Electra and PMMA films, respectively. $500\,$nm for PMMA thickness was found to be optimal value for successful transfer. Twice thinner films were found to be excessively fragile, whereas twice thicker films were loosing the flexibility. In the step 2 the lithography of the required structure is performed, which is followed by the development of the exposed regions of the PMMA film in the MIBK/IPA (1:3) developing solution. It is important to note here that Electra was found to be almost insoluble by neither IPA, nor MIBK, nor their mixture. In the step 3, the sacrificial substrate with Electra layer and developed PMMA is placed in water. PMMA is hydrophobic, which allows the floating of the whole substrate on the water surface. Electra is gradually dissolved by water which results in the separation of the PMMA film from the substrate. The substrates sinks to the bottom of the water container, whereas PMMA stays floating. 

In step 4 PMMA film is picked up from the water surface by an appropriately designed carrier. Using a transfer stage \cite{transferstage} the carrier with the PMMA film is aligned with the target material/sample, on the surface of which it is subsequently released (the result is illustrated as step 5 in Fig. 2(a)). Typically, such release process implies melting of the transferred film in order to detach it from the carrier/stamp \cite{Zomer2014}. However, in our case melting of the PMMA would result in substantial degradation of the structure of the developed PMMA regions, thus, it has to be done at temperatures below the melting point of PMMA ($137^{\circ}$C). Detailed description of the transfer and low temperature release of the PMMA film from the carrier is given in the supplementary materials. Directly after step 5 the sample is placed in the film deposition equipment, where the required (metallic) films are formed. Subsequently, the PMMA mask can be removed by a common lift-off process in acetone, resulting in the final structure on the surface of the target sample illustrated as step 6.

In the whole procedure described above PMMA is never melted on the surface of the target sample, which implies that it is not strongly attached to it. This means that PMMA mask after the metal deposition can be detached mechanically. Such detachment allows to avoid dipping of the sample in the solvent together with the PMMA mask, and therefore protects the sample surface from possible PMMA contamination. Nevertheless, we expect that the contamination occurs mostly when PMMA is fully melted on the surface, since large contamination particles become mobile within PMMA layer and can attach to the sample surface (as it happens for Fig.~1(a,b)). Therefore, doing the lift off in acetone at step 6 is not expected to contaminate the sample surface considerably. This assumption has not been conclusively confirmed in our experiments and for most of the fabricated samples mentioned in this work we used mechanical detachment procedure that is described in the supplementary materials.

It is important to indicate the limitations of this method. Firstly, since the PMMA mask is prepared via the e-beam lithography in a $500\,$nm PMMA film, it already implies that making any openings, that are of the order of $\simeq100\,$nm or smaller, is challenging. Secondly, during the pick up of the PMMA from water and its transfer to the targeted substrate, the whole PMMA structure may get distorted to a certain extent depending on the design of the openings and on the appropriate care taken during the transfer/placement process. Nevertheless, we believe that the structure size of $~300\,$nm should be achieved quite reliably as it is confirmed by our experience. Fig.~1(d) shows an example of the PMMA mask where the arms of the Hall bar were designed to be $500\,$nm wide. Furthermore, one should note that the non-rigid PMMA film with openings has to sustain the suspension in the air, which implies, for example, that extended and curved openings might cause excessive sagging of the PMMA parts and result in distortions incompatible with the targeted design. Nevertheless, any ultimately complex design of the target structure of the electrodes can be made in principle by splitting it into two (or more) simpler structures prepared separately similarly to the case shown in section~4.

Similar methods for the fabrication of the contamination-free contact interfaces using PMMA as a mask have been reported in \cite{Cai2018}, where the authors used KOH at 95$^{\circ}$C in order to detach the pre-patterned PMMA film from the Si/SiO$_2$ substrate. In general this method provides very similar resolution and flexibility, only that in our case the sacrificial Electra layer can be dissolved in water at room temperature conditions which simplifies the fabrication procedure compared to the method developed in \cite{Cai2018}.

\begin{figure*}[!]
\centering
\includegraphics[width=\linewidth]{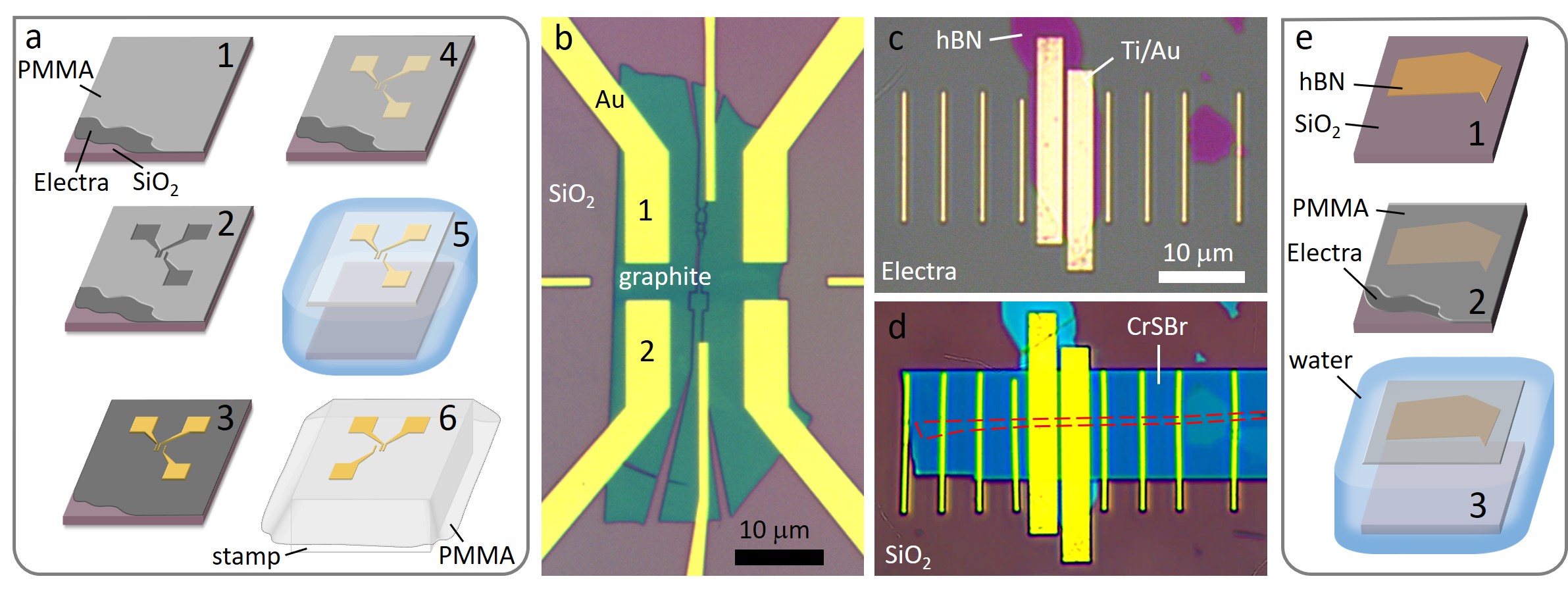}
\caption{(a) Schematically shown recipe steps for the preparation of the PMMA membrane with embedded metallic electrodes. (b) Graphite flake that was contacted by the electrodes prepared by the recipe shown in panel (a). Panels (c) and (d) show optical images of the steps of preparation of Ti/Au/hBN/graphene/CrSBr device. Image in panel (c) corresponds to the step 3 in panel (a). Image in panel (d) correspond to the fabrication stage after the step 6. (e) Recipe steps for an alternative pick up method of the layered materials from the surface of the Si/SiO$_2$ substrate. It is suitable only for materials that do not degrade when in contact with water such as hBN or graphene.}
\label{fig2}
\end{figure*}

\subsection*{Examples of the recipe usage}
Using the method described in this section it is possible to fully exclude the presence of PMMA contamination (such as that shown in Fig.~1(a,b)) from the interface between the (metallic) electrodes and the studied sample/material. This technique was successfully employed in the fabrication of the spin-orbit torque devices, where it was demonstrated that a pristine interface between the WSe$_2$ flakes and deposited Py is essential for observing a clear interfacial spin-orbit torque signals \cite{Hidding2021}. Furthermore, in \cite{Damerio2021} we employed this technique for fabricating Pt Hall bars on top of the antiferromagnet CaFe$_2$O$_4$, where we observed spin Hall magnetoresistance that crucially depends on the quality of the interface between Pt and CaFe$_2$O$_4$. The images of one of the Pt/CaFe$_2$O$_4$ structures is shown in Fig.~2(d) and (e) corresponding to the steps 5 and 6 of panel (c), respectively. Finally, the PMMA films can be used not only as masks for the metal deposition, but also as masks for shaping of the targeted sample by means of milling/reactive-ion etching. Fig.~2(f) demonstrates an optical image of a monolayer graphene flake that was etched through the PMMA mask prepared with the described above method. After the etching the PMMA mask was removed mechanically, leaving a Hall bar shaped graphene flake leaving minimal amount of contamination on the surface. 

Preserving the pristine interface between the targeted material and deposited films is often essential for the successful realisation of the designed sample structure/experiment. The fabrication method described in this sections provides relatively simple means for achieving such pristine interfaces and can be easily adapted for a variety of research areas. Moreover, the flexibility of the PMMA film allows preparation of the metallic films on curved surfaces such as fabrication of metallic structures on the surface of the optical fibers/lenses as described in \cite{Cai2018}.

\section{Damage-free contacts for air-sensitive and fragile materials}

Commonly used procedure for preparing the electrodes (based on lithography with subsequent deposition of metallic contacts) can be used for a wide range of solid state materials. However, there is a significant portion of materials that cannot withstand the conditions of such procedure. For example, air sensitive materials has to be kept under inert atmosphere to avoid degradation, which implies considerable limitations on the fabrication process. Moreover, organic materials, that are nowadays challenged as a platform for the functional solid state devices, are very fragile and usually degrade when either immersed in the solvent (such as acetone), or put under vacuum, or heated up. For such materials the regular lithography based electrode preparation procedure simply cannot be used. Furthermore, commonly used deposition/evaporation of metallic films can damage the surface of the studied semiconductor material, thus triggering the formation of the chemical disorder and Fermi-level pinning at the contact interface which results in formation of the unwanted Schottky barriers.

In this section we offer a viable solution for preparing the electrodes for air-sensitive and fragile materials. Essentially, we developed the recipe where the electrodes are prepared on a separate substrate and later are transferred to the target material under inert atmosphere at room temperature. This method implies that at all stages the target materials is never immersed in the solvent and can always be kept at room temperature and under inert atmosphere. 

\subsection*{The recipe}
Fig.~2(a) summarises the steps of the developed fabrication method. Steps 1 and 2 are exactly the same as those described in Fig.~1(c), where we spin-coat the sacrificial Si/SiO$_2$ substrate with Electra and PMMA layers and later expose the PMMA layer with e-beam, such that the shape of the electrodes are formed in the PMMA film after development. Then, the Si/SiO$_2$ substrate is loaded into a deposition system and metallic contacts of desired composition are formed. Electra film was found to be able to withstand dipping in acetone thus allowing regular lift off. Deposited metallic contacts on the Electra film after lift off are schematically shown in the schematics of the step 3. In step 4 the prepared structure is spin-coated with A4 PMMA at the rate of $1000\,$rpm and baked at $180^{\circ}$C for $90\,$sec. In step 5 the substrate is immersed in water, where the Electra film gets dissolved and PMMA film (containing the metallic electrodes) is separated from the Si/SiO$_2$ substrate. Then, the PMMA film is picked up from the water surface and placed on the polydimethylsiloxane (PDMS) stamp that is separately prepared on the glass slide. Before placing PMMA film onto the PDMS stamp, it is flipped over such that the electrodes' surface is facing outwards (see supplementary material for more details). In the next step, using the transfer stage, the PMMA film with the electrodes is placed on top of the target material.

\subsection*{Examples of the recipe usage}
In Fig.~2(b) we show the graphite flake for which the Au electrodes were made with the help of the procedure described above. The 2 terminal resistance between contacts 1 and 2 was found to be $\simeq440\,\Omega$ which implied that contact resistance between the transferred Au electrodes and graphite is lower than $\simeq220\,\Omega$ suggesting a reasonable Ohmic contact at the interface.

Furthermore, the fabrication and transfer of the electrodes can be combined with stacking of layered materials. Fig.~2(c) gives an optical image of the Ti/Au electrodes prepared on the Electra film. It corresponds to step 3 of panel (a) with the difference that prior the fabrication of the electrodes we additionally exfoliated hexagonal boron nitride (hBN) flake on the surface of Electra. As it is seen, two of the Ti/Au electrodes cover the magenta-coloured hBN flake. After spin-coating the sample with PMMA (step 4), it was detached from the sacrificial substrate via dissolving Electra in water (step 5), was attached to the PDMS stamp (step 6) and then was used to pick up a graphene flake from a separate SiO$_2$/Si substrate. Subsequently, the full stack of Au/Ti/hBN/graphene was placed on top of another layered material CrSBr. The stack in its final configuration on top of the SiO$_2$/Si substrate is shown in panel (d), still covered with PMMA film. The sample was further processed for the lithography and deposition of an additional layer of ferromagnetic electrodes and resulted in the experimental work published in \cite{Kaverzin_2022}. During the full fabrication procedure of stacking and preparation of 2 different types of electrodes the graphene surface was covered by the polymer (PMMA) only once, thus minimising the level of the surface contamination.

Fabrication methods of the metallic electrodes on the sacrificial substrate with the subsequent transfer of them on the target material have been reported elsewhere \cite{Liu2018,Telford2018,Jung2019,Makita2020,Song2021,Wang2022,Liu2022,Zhang2023}. The main difference of our method compared to the previous reports is the employment of the water soluble Electra layer, which can easily withstand solvents and ensures an easy and smooth detachment of the PMMA layer with the electrodes from the sacrificial substrate.

\subsection*{Alternative method for picking up the 2D materials}
In addition, we would like to mention that the exfoliation of 2D materials (hBN in the case of the sample shown in Fig.~2(c)) on the surface of the Electra film from the scotch tape was found to be rather difficult and unreliable. From our experience, in many occasions the Electra film had stronger interaction with the scotch tape rather then with the Si/SiO$_2$ substrate, such that after removing the tape the Electra film got picked up and stayed on the tape. In order to overcome this issue we offer a more reliable method for the preparation of the hBN flakes on the surface of Electra. The procedure is summarised schematically in Fig.~2(e) and in essence offers the recipe for the picking up of the layered materials from the surface of the Si/SiO$_2$ substrate by the polymer film.

In step 1 we exfoliate hBN on the surface of the Si/SiO$_2$ substrate. In step 2 we spin-coat Electra and PMMA films using the same parameters as described in section 1. In step 3 we place the substrate in water where the Electra film gets dissolved, however, during the detachment of the PMMA film from the substrate the hBN flakes usually get detached from the surface of the SiO$_2$ and get attached to the PMMA surface. The exact scenario of the interaction between water, SiO$_2$, PMMA and hBN is unclear, yet it might be related to the fact that both PMMA and hBN are hydrophobic while SiO$_2$ surface is hydrophilic after spin-coating with Electra and dipping it into water. After picking up the PMMA film from water with hBN flakes attached to it, hBN can be transferred to any other substrate including the preliminary prepared Si/SiO$_2$ substrate spin-coated with Electra. After releasing PMMA on the Electra surface it is possible to perform e-beam lithography for the preparation of the Ti/Au electrodes and gates in order to obtain the structure similar to that shown in Fig.~2(c).

\begin{figure*}[ht]
\centering
\includegraphics[width=\linewidth]{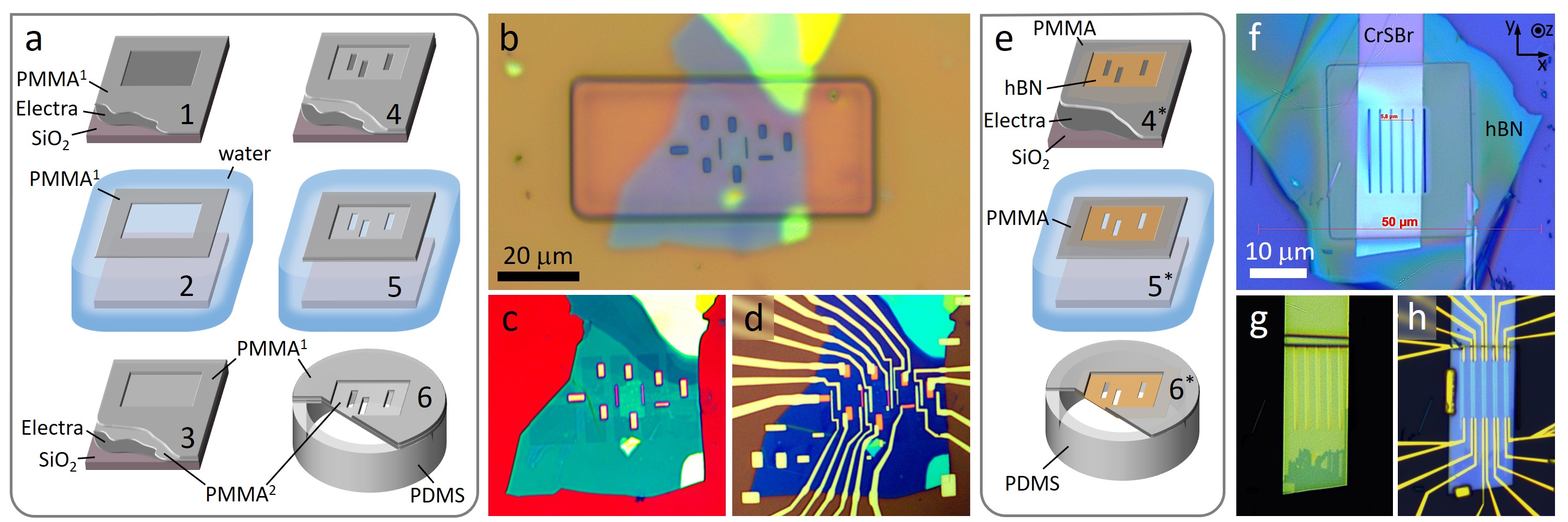}
\caption{(a) Schematically shown recipe steps for the preparation of the double layer PMMA shadow masks. (b) Optical image of a double layer PMMA shadow mask placed on top of the sample. Closest to the optical lens PMMA$^2$ layer is in focus while the sample itself is out of focus. Panels (c) and (d) show the same device as in (b) but after Ti/Au deposition (c) and after TiOx/Co deposition (d). (e) An alternative recipe where instead of layer PMMA$^2$ we use hBN that was preliminary milled by FIB. Panel (f) shows an example of using hBN/PMMA mask placed over the CrSBr flake. Optical images of the same device made after deposition of Pt electrodes (g) that are further connected by Ti/Au (h).}
\label{fig3}
\end{figure*}

We would like to note that in both cases shown in panels (b) and (d) of Fig.~2 we released the PMMA film with electrodes on the targeted substrate by increasing the temperature beyond the melting point of the PMMA ($\simeq137^{\circ}$C). However, the room temperature release of the PMMA film is also possible (see detailed description in the supplementary material). Furthermore, it is nowadays common to install the transfer stages inside a glove box where the atmosphere is inert. Therefore, the final stage when the electrodes are placed on top of the sample can readily be done in the controlled atmosphere, thus allowing the preparation of the contacts of fragile and/or air-sensitive materials using the recipe described above. 

\section{Nanoscale shadow masks}

In section~1 we described the recipe that allows fabrication of a contamination free interface between the electrodes and the studied material. Yet the surface outside of the electrodes' regions still gets in direct contact with the PMMA film and, thus, can get contaminated by it or other residues (Electra residues, moisture, etc.), especially if the PMMA mask is removed by dissolving it in acetone. On the other hand, if the PMMA mask is removed mechanically, there is a finite probability that PMMA film will pick up the targeted flakes (or parts of them). This is in particular likely if the sample (together with the PMMA mask) is heated during the metal deposition which promotes the adhesion of PMMA to the flake. In this section we describe a recipe, which allows to fully avoid the contact of the studied materials with any of the polymers, thus excluding both the possibility of contamination and unwanted pick up of the flakes. This recipe might be primarily useful for the cases when the sample surface has to be preserved as clean as possible, for example for the scanning tunneling microscopy, where even low contamination level of the surface prohibits the observation of the crystal structure and/or spectroscopy.

\subsection*{The recipe}
Fig.~3(a) shows schematically the recipe steps. This recipe requires two distinct PMMA layers, which we will call PMMA$^1$ and PMMA$^2$. Preliminarily, we spin-coat two separate sacrificial SiO$_2$ substrates with Electra and PMMA films using the same parameters as described in section~1. At the step 1 we perform an e-beam lithography on the substrate with PMMA$^1$ in order to make a rectangular opening with the typical dimensions of $30\,\mu$m by $70\,\mu$m. Generally we make a 10 by 10 array of such openings separated by $\simeq300\mu$m, yet only one of such openings is shown in the schematics. As we will see later PMMA$^1$ serves only as a spacer which separates the target material from the actual shadow mask. This implies that the size of the opening in it has to be bigger than the size of the sample area that is targeted to be preserved intact (typically bigger than the size of the flake of the studied 2D material). Further, we place the substrate with structured PMMA$^1$ into water where the Electra layer gets dissolved resulting in a floating PMMA$^1$ on the water surface, step 2. Subsequently, the other sacrificial substrate with so far unexposed PMMA$^2$ is dipped into the same water container and is used to pick up PMMA$^1$ such that we end up with a stack Si/SiO$_2$/Electra/PMMA$^2$/PMMA$^1$ (from bottom to top), shown as step 3. Next, using e-beam lithography we make the designed openings in PMMA$^2$ (step 4) which will be used as slits for metal deposition at a later stage. The resulting structure is placed in a fresh water container where the sacrificial substrate sinks to the bottom leaving floating double layer PMMA$^2$/PMMA$^1$, step 5. In step 6 we pick up the PMMA double layer and place it across a PDMS stamp that has an opening in the middle such that PMMA$^1$ is facing up. Typically, PDMS stamps used for the transfer of 2D materials have rectangular shape and do not have any openings/voids, however, the prepared PMMA double layer with small openings in PMMA$^2$ is rather fragile and therefore the direct contact with PDMS should be avoided. In the following steps the prepared stamp is aligned with the studied flake and placed on the surface of the targeted substrate such that PMMA$^1$ is on contact with the substrate and PMMA$^1$ is suspended over the targeted flake. After a gentle mechanical release of the PMMA stack from the PDMS stamp the sample is placed into a deposition system, where the required metals are evaporated through the slits in the PMMA layer. Finally, the PMMA mask is detached mechanically from the surface of the sample. The image of the used PDMS stamp and details on its preparation/handling are given in the supplementary material.

\subsection*{Examples of the recipe usage}
In Fig.~3(b) we show an optical image of the device covered by the PMMA double layer. The bottom layer of PMMA (PMMA$^1$) is in direct contact with the substrate everywhere except the central rectangular shaped region. The top PMMA$^2$ layer is separated from the sample surface by the thickness of the bottom PMMA$^1$ (which is about $500\,$nm and has openings of the desired shape (in focus of the image) aligned with the targeted structure (out of focus). The image of the sample, after deposition of Ti/Au electrodes and mechanical removal of the PMMA double layer, is shown in panel (c). The size of the narrowest Ti/Au electrodes prepared on this device was about $300\,$nm. The sample was further processed to prepare another set of (ferromagnetic) electrodes, panel (d), and was used for the spin transport measurements published in \cite{Ghiasi2019}. Efficiency of the spin-sensitive electrodes crucially depends on the contamination of their interface with the studied material. For the case when 2 different types of the electrodes are required the described above recipe allows to decrease the number of the performed regular lithography steps down to one, thus minimising the level of the contamination of the electrodes interfaces.
 
The idea of deposition of metallic electrodes on the surface of the targeted sample through the shadow masks is not new and has been employed in the literature for the preparation of the solid state devices \cite{Deshmukh1999,Bao2010,Shih2014,Amato2018,Sun2023,Han2023}. Such shadow masks are commonly called stencil masks and are made out of a solid material such as low tension silicone nitride. The distance between such rigid macroscopic-sized masks and the target substrate is hard to control precisely and to make it as small as $\simeq1\,\mu$m (although not impossible \cite{Deshmukh1999}). Commonly, the distance between the stencil mask and the substrate is above $10\,\mu$m. This substantially reduces the resolution of the deposited structures, unless the beam of the depositing particles is collimated which unfavourably reduced the deposition rate to a large extent. Instead, our recipe allows the preparation of easy-to-prepare PMMA shadow masks, flexibility of which ensures the conformity of the mask to the local sample topography and thus minimises the gap between the openings in the mask and the targeted substrate ($<1\,\mu$m).

\subsection*{FIB-ed hBN as a stable shadow mask}

\begin{figure*}[ht]
\centering
\includegraphics[width=\linewidth]{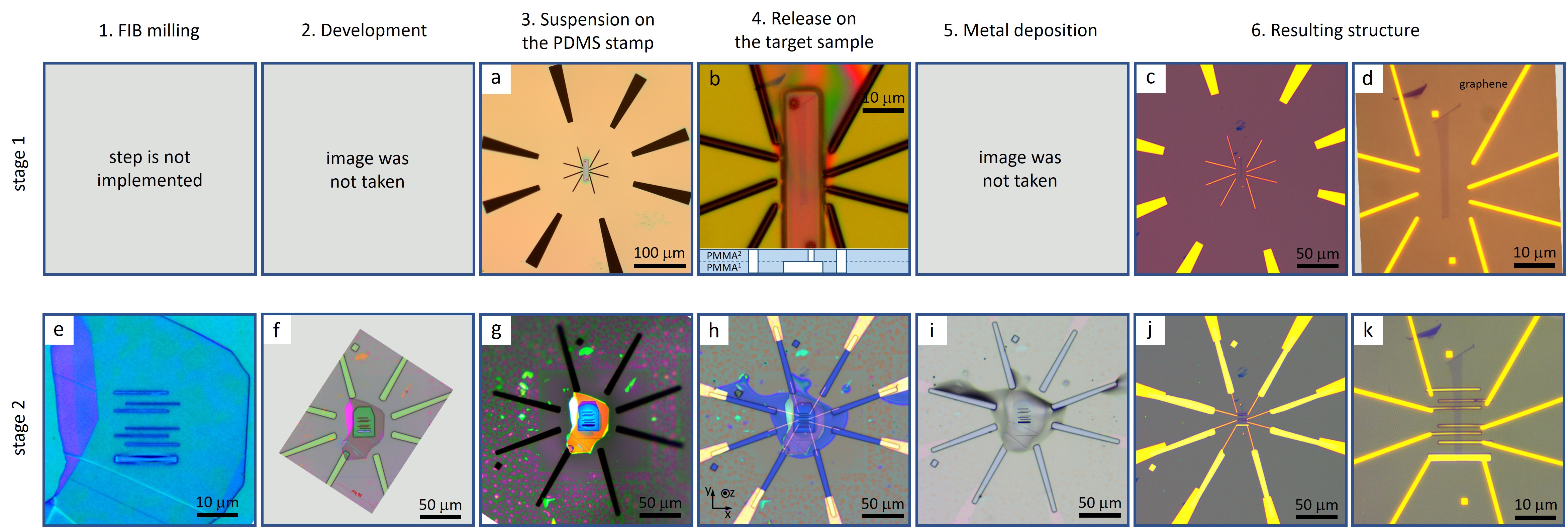}
\caption{Optical images of the steps performed for the fabrication of the graphene device with TiOx/Co electrodes using residue free hBN/PMMA shadow masks. Top and bottom rows of the images correspond to the stage 1 and stage 2 of the fabrication process. The "columns" are labeled corresponding to the fabrication step performed directly before taking the image. The steps where the images were not taken are designated with "image was not taken". Definitions of PMMA$^1$ and PMMA$^2$ used here in panel (b) are given in Fig.~3(a).}
\label{fig4}
\end{figure*}

The described above recipe has two relevant limitations. Firstly, the openings in the layer PMMA$^2$ cannot be placed too close to each other. From our experience we found that any separation smaller than $\simeq500\,$nm would result in an unstable structure. During the placement of the shadow mask on the targeted devices tension and possible wrinkles in the suspended PMMA double layer can result in a substantial distortion of the openings with breaking of the narrow parts separating the neighbouring openings. Secondly, depending on type and dimensions of the used equipment, during the metal deposition the sample together with the PMMA shadow mask can heat up which is likely to result in a continuous sagging of the PMMA and distortion of the openings. In addition to that, deposited metal on the surface of the shadow mask contributes to the tension in the mask and causes additional distortion. We commonly use e-beam evaporation method for metal deposition. To minimize the distortion effects we limit the thickness of the deposited metal by $\simeq30\,$nm (which is typically sufficient to provide reasonable conductivity of the electrodes) and use moderate deposition rates of about $0.5\,A$/s. With these parameters the distortion in the shape of the deposited electrodes was usually less than $100\,$nm that is acceptable for most of our structures.

The stability of the shadow mask can be substantially improved by replacing the PMMA$^2$ layer with hBN that we found to be more enduring against heat and stress. In Fig.~3(e) we show schematically the main steps necessary for preparing the shadow mask made from hBN. At first, we prepare hBN flakes on the surface of Electra, which is spin-coated in advance on top of the Si/SiO$_2$ substrate. This can either be done by the exfoliation of hBN directly on the Electra surface or via the method described in Fig.~2(e). Next, we find optically suitable hBN candidates that are of moderate thickness (about $40\,$nm) and load the substrate with hBN flakes into the focused ion beam setup. Using the beam of Ga ions we mill the openings in the targeted hBN flakes. Directly after that the substrate is spin-coated with A4 PMMA at $1000\,$rpm, baked for $90\,$sec at $180^{\circ}$C and processed with e-beam lithography in order to make rectangular-shaped openings in the PMMA layer directly on top of the milled parts of the hBN flake, resulting in a structure shown in step 4* of the panel (e). Note here that the opening in PMMA has to be within the area of the hBN flake. In step 5* we put the substrate into water which results in a floating hBN/PMMA structure. Subsequently, the floating mask is picked up from water and placed on the PDMS stamp with an opening in the center, step 6*. After the preparation such shadow mask can be used in exactly the same fashion as that described in Fig.~3(a). 
 
We found that hBN is substantially stronger than PMMA and therefore can withstand multiple metal depositions. This, in turn, implies that we can perform the metal deposition at two distinct angles with respect to the out-of-the-sample-plane axis, which results in electrodes that are placed with a distance of $\simeq100\,$nm or even smaller from each other. In Fig.~3(f) we show an optical image of the CrSBr flake (preliminarily exfoliated on the surface of the Si/SiO$_2$ substrate) covered by the appropriate hBN/PMMA shadow mask. This sample was loaded into an e-beam metal evaporation setup. We deposited $15\,$nm of Pt at the angle of $+30\,$degrees between z-axis of the sample and the direction of the metal evaporation (axes are labeled in panel (f)) and then another $15\,$nm of Pt at the angle of $-30\,$degrees. Assuming $500\,$nm separation between the hBN and the surface of the sample, and $300\,$nm width of the slits, we obtain $\simeq250\,$nm separation between the two sets of the deposited Pt electrodes which can be approximately confirmed from the optical image of the device after the shadow mask was mechanically removed, Fig.~3(g). The sample was further processed with a regular e-beam lithography in order to prepare the Ti/Au connections to the Pt electrodes, Fig.~3(h). This Pt/CrSBr sample was aimed to study the magnon transport in an insulating CrSBr, where the quality of the interface between Pt and CrSBr and short distance between the neighbouring Pt electrodes were essential for a successful observation of magnon-related signals \cite{magnonCrSBr}. In principle we could use the recipe described in section~1 for making a pristine contact interface, however, within that recipe it is very challenging to place electrodes at distances below $500\,$nm, whereas we wanted to position the electrodes as close as possible for maximising the size of the detected non-local signal.

\section{Fabrication example:\\ contamination-free deposition of sub-$\mu$m scale electrodes}

In this section we describe a complete fabrication sequence for preparation of a specific graphene device, where the graphene flake is never in contact with any of the polymers and is never intentionally heated above 70$^\circ$C (the temperature of the device during the metal deposition is not controlled). The fabrication procedure generally follows the recipe discussed in section~3 and offers practical details of the implementation of the earlier described steps. The full procedure is divided into two stages. The main reason here is that it is close to impossible to make a suspended PMMA mask for a full length of macro-sized contacts due to accumulation of the excessive tension near the central part of the PMMA due to sagging. Such excessive tension is very likely to results in a breaking of the central part of the suspended PMMA during any of the steps 3-6 from Fig.~1(c) or 5-6 from Fig.~3(a). Therefore, all the electrodes are split into two discontinuous but overlapping sections (Fig.~4) which are fabricated consecutively and independently from each other. 

\subsection*{Stage 1}

\begin{figure}[ht]
\centering
\includegraphics[width=\linewidth]{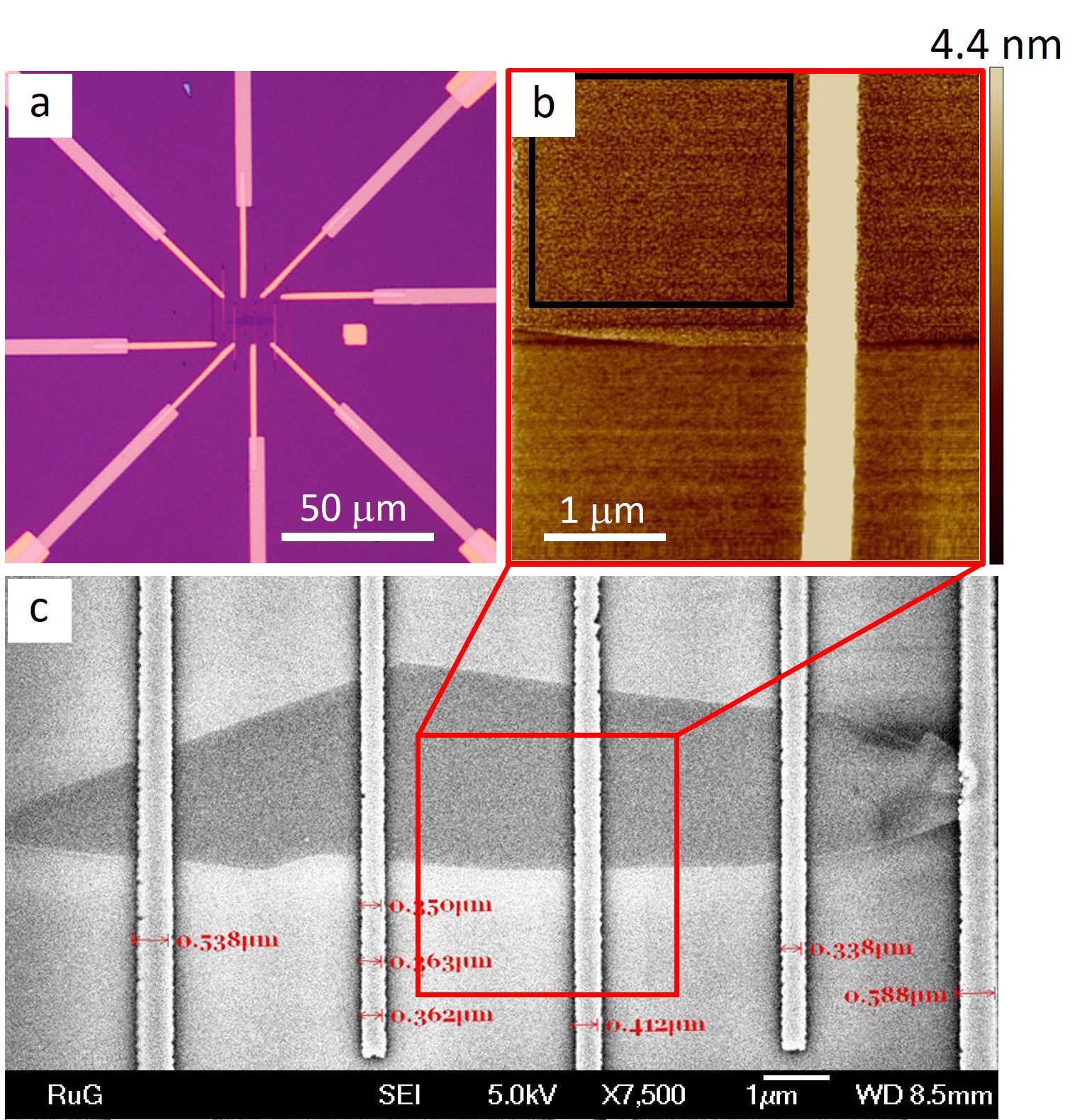}
\caption{(a) Optical image of the graphene device prepared with the help of the double layer PMMA shadow mask. (b) AFM scan of the same device. The scanned area is marked with red square in the SEM image of the devices shown in panel (c).}
\label{fig3}
\end{figure}

Top set of panels (a)-(d) in Fig.~4 corresponds to the stage 1 of the device fabrication. The sections of the electrodes prepared during the stage 1 can be seen in panel (a) where the optical image of double layer PMMA mask attached to the PDMS stamp is shown (equivalent to the schematics of step 6 in Fig.~3(a)). As one can see the structure consists of disconnected openings that are distributed evenly over the area in order to decrease the stress and sagging in the suspended PMMA membrane. The openings correspond to the linking parts of the electrodes and are not directly connected to the graphene flake. The central part of the mask, which is later aligned to be on top of the graphene flake, has no openings except the two small ones required for pressure equalization. In panel (b) we show an optical image of the device after the PMMA mask is placed on the surface of the substrate with graphene. The graphene gets enclosed within a $40\times10\times0.5\,\mu$m$\times\mu$m$\times\mu$m compartment within the PMMA film which protects the flake during the metal deposition while keeping it intact. Such compartment is made from two distinct PMMA layers following the recipe from Fig.~3(a). For clarity at the bottom of the panel (b) we schematically show the cross section of the PMMA mask corresponding to the bottom edge of the optical image. After depositing Ti/Au films with corresponding thicknesses of $\simeq5/25\,$nm we mechanically remove the PMMA mask (supplementary material). At the and of stage 1 we get a sample with deposited Ti/Au links as shown in panels (c) and (d).

\subsection*{Stage 2}

In stage 2 the complementary parts of the metallic electrodes are made. In order to minimise the widening of the slits in the mask during the metal deposition we employed the hBN-based recipe, Fig.~3(e). At first, we exfoliated hBN flakes on the surface of the Electra, preliminarily spin-coated on a separate Si/SiO$_2$ substrate. Then, selected hBN flake with appropriate thickness ($\simeq40\,$nm) was loaded into a focused ion beam system, where the openings were milled with the help of the Ga$+$ beam, panel (e). Next, the substrate with milled hBN flake is spin-coated with 950K A4 PMMA, which is followed by the e-beam lithography. The subsequently developed structure is shown in panel (f). The mask suspended over the PDMS stamp is shown in panel (g). The stamp with the mask was aligned with the targeted structure such that the openings connect the already deposited Ti/Au parts together and with the graphene flake, thus making continuous electrodes. Released on the sample surface mask is shown in panel (h) where one can notice that it does not lie flat, possibly due to the additional stress caused by the presence of the hBN flake. This evidently results in an increased distance between the targeted graphene flake and the shadow mask, which instead of $\simeq500\,$nm becomes larger than $1\,\mu$m. In most of the cases this is still sufficiently close to the sample surface in order to have reasonably well defined shape of the electrodes, yet for the most optimal performance one should aim to make the mask as flat as possible. Examples of good sitting masks are shown in Fig.~3(b),(f) and in Fig.~4(b). 

As it is also noticeable from Fig.~4(h) the final alignment of the PMMA mask with the Ti/Au parts was not optimal. The openings in hBN are shifted along y-axis by $\simeq1\,\mu$m with respect to their targeted locations. In order to compensate for this shift, the metal deposition was performed under an angle of $\simeq30\,$degrees with respect to the z-axis within the z-y plane. The optical image of the sample after deposition of the Al$_2$O$_3$/Co films of $1/30\,$nm thicknesses (the sample was aimed for performing the spin transport and thus required spin-sensitive electrodes) is shown in panel (i). Finally, we performed mechanical removal of the PMMA mask from the sample surface. In panels (j) and (k) we show the final structure of the metallic electrodes that consist from Ti/Au and Al$_2$O$_3$/Co parts.

\subsection*{Characterisation of the contamination level}

Using the method described above we have prepared $\simeq10$ different graphene devices, for most of which we used PMMA masks with no hBN inclusion, thus following the recipe Fig.~3(a) exactly. In order to characterise the cleanliness of one of the devices prepared in such way (its optical image is shown in Fig.~5(a)) we performed an AFM scan of the sample surface, panel (b), and SEM imaging, panel(c). The area over which the AFM was performed is marked in the SEM image with the red square. As one can see there is no polymer-like contamination (as visible in Fig.~1(b)) detected neither by AFM or SEM. Quantitatively this is characterised with the roughness of the sample which was found to be $\simeq0.4\,$nm within the area outlined as black rectangular in panel (b). Such value is typical for the roughness of the pristine Si/SiO$_2$ substrate and thus indicate absence of any contamination on the graphene surface.

\section{Conclusions}
We have developed and successfully employed the fabrication recipes based on flexible PMMA films/shadow masks, that are aimed at making metal connections to the layered materials with minimising the influence of fabrication processing on the quality of the final device. Namely, in section~1 we described the recipe for preparing a residue-free electrode interface between the deposited metal and the studied material. Such pristine contact is essential, for example, for an efficient spin transport across the interface as was demonstrated in \cite{Hidding2021}. In section~3 we described the recipe for preparing the shadow masks that can be placed at a very close vicinity to the sample surface (as close as $500\,$nm), thus allowing the preparation of narrow contacts ($\simeq300\,$nm wide) with relatively flexible design an accuracy of the e-beam lithography. Using this recipe the direct contact between the studied materials and used polymers can be fully excluded allowing keeping the sample at a pristine state. This recipe was further improved by replacing one of the mask layers by an hBN flake which improves the stability of the mask and allows multiple tilted depositions. It further broadens the limitations of the contamination-free electrode preparation technique with separation between the contacts below $100\,$nm. Finally. in section 2 we offer the recipe for the fabrication of the electrodes for the fragile and/or heat/oxygen/moisture sensitive materials. All the described methods were employed for the fabrication of the devices that were subsequently used for producing published experimental results. The developed recipes are all based on using easy-to-handle PMMA and Electra polymers and have a broad application potential for various areas of the fundamental research where the quality of the material and that of the interfaces with the electrodes plays an essential role and limits the range of the accessible physics phenomena.

\section*{Acknowledgements}

The authors would like to thank Josep Ingla-Aynés, Talieh S. Ghiasi and Takashi Kikkawa for fruitful discussions on various fabrication methods and for collaborations resulted in the publications mentioned in this manuscript. We thankfully acknowledge the technical assistance from Johan Holstein, Martijn de Roos, Tom Schouten and Hans de Vries. We also owe a kind gratitude to Geert ten Brink and Prof. Bart Kooi for allowing to use the focused ion beam system and giving the detailed instructions on its usage. The presented research was funded by the Dutch Foundation for Fundamental Research on Matter (FOM) as a part of the Netherlands Organisation for Scientific Research (NWO), the European Union's Horizon 2020 research and innovation program under grant agreements No 696656 and 785219 (Graphene Flagship Core 2 and Core 3), the Zernike Institute for Advanced Materials, and the Spinoza Prize awarded in 2016 to B. J. van Wees by NWO. The performed research was also supported by JST-CREST (JPMJCR20C1 and JPMJCR20T2), Grant-in-Aid for Scientific Research (JP19H05600) and Grant-in-Aid for Transformative Research Areas (JP22H05114) from JSPS KAKENHI, Japan, and Institute for AI and Beyond of the University of Tokyo.


\begin{thebibliography}{10}
\bibitem{transferstage}
The transfer stage was purchased from HQ Graphene.
\end{thebibliography}
\end{document}